\documentclass[12pt]{iopart}

\usepackage[dvips]{graphicx}
\usepackage{array}
\usepackage{amssymb}
\usepackage{hhline}
\usepackage{longtable}
\usepackage{dcolumn}
\usepackage{bm}
\usepackage{subfigure}
\usepackage[hyperindex,CJKbookmarks,dvipdfm,colorlinks,citecolor=blue,linkcolor=blue]{hyperref}
\begin{document}

\title{Epidemic spreading with nonlinear infectivity in weighted scale-free networks}

\author{Xiangwei Chu$^1$, Zhongzhi Zhang$^2$${}^,$$^3$, Jihong Guan$^1$, and Shuigeng
Zhou$^2$${}^,$$^3$}

\address{$^1$School of Electronics and Information, Tongji University,
4800 Cao'an Road,Shanghai 201804, China}
\address{$^2$School of Computer Science, Fudan
University, Shanghai 200433, China}
\address{$^3$Shanghai Key Lab of
Intelligent Information Processing, Fudan University, Shanghai
200433, China}

\eads{zhangzz@fudan.edu.cn, jhguan@tongji.edu.cn,
sgzhou@fudan.edu.cn}

\begin{abstract}
In this paper, we investigate the epidemic spreading for SIR model
in weighted scale-free networks with nonlinear infectivity, where
the transmission rate in our analytical model is weighted.
Concretely, we introduce the infectivity exponent $\alpha$ and the
weight exponent $\beta$ into the analytical SIR model, then examine
the combination effects of $\alpha$ and $\beta$ on the epidemic
threshold and phase transition. We show that one can adjust the
values of $\alpha$ and $\beta$ to rebuild the epidemic threshold to
a finite value, and it is observed that the steady epidemic
prevalence $R$ grows in an exponential form in the early stage, then
follows hierarchical dynamics. Furthermore, we find $\alpha$ is more
sensitive than $\beta$ in the transformation of the epidemic
threshold and epidemic prevalence, which might deliver some useful
information or new insights in the epidemic spreading and the
correlative immunization schemes.
\end{abstract}

\maketitle

\section{Introduction}
There has been a long history for the research of epidemic
spreading~\cite{KM1927}. And in the general case, the epidemic
system can be represented as a network where nodes stand for
individuals and an edge connecting two nodes denotes the interaction
between individuals. In the past, researchers mainly focused the
disease transmission study on the conventional
networks~\cite{Bailey,May1992} such as lattices, regular tree, and
ER random graph. Since late 1990s, scientists have presented a
series of statistical complex topological
characteristics~\cite{RAAB,LANAmaral,SNDJEFM03,PVCambridge} such as
the small-world (SM) phenomenon~\cite{SM98} and scale-free (SF)
property~\cite{BA99} by investigating many real networks including
the internet~\cite{Faloutsos}, the www~\cite{AlbertJeong}, the
scientific web~\cite{NewmanSCN01}, the protein
networks~\cite{protein} and so on. Subsequently, the studies of
dynamical processes on complex networks also have attracted lots of
interests with various subjects~\cite{NewmanRev03,SVYMRev06}, and as
one of the typical dynamical processes built on complex networks,
epidemic spreading has been investigating intensively once more.

The basic conceptual tools in understanding the epidemic spreading
and the related effective strategies for epidemic controlling should
be the epidemiological models~\cite{Daily01,Diekmann}. Among the
numerous possible models, the most investigated and classical models
are SI model~\cite{BathelemySI04,BathelemySI05}, SIS
model~\cite{PVSIS01,PVSIS02,BogunaSIS02,BogunaSIS03,yangangepjb08}
and SIR model~\cite{RMMAYSIR01,LloydSCIENCE,MPV02,Moreno03}, which
can approximately describe the spreading of real viruses such as
HIV, encephalitis, influenza virus in biological networks; computer
virus, trash mail in technological networks; and even gossip in
social networks. The most valuable result in standard SIR model (or
SIS model) is that the critical threshold (of transmission rate)
vanishes for the scale-free networks in the limit of infinite
network size~\cite{MPV02}.

In consideration of epidemic spreading in real cases, there yet has
been some inappropriate assumption in the details of the standard
SIR model. We know that in the classical SIR model the transmission
rate $\lambda$ is a constant, but in the real world, $\lambda$
should be different among individuals. Based on this assumption, in
reference~\cite{JooLeb04}, Jaewook Joo et al. proposed the effective
transmission rate that they introduced a effective coefficient
$C(k,l)$ based on standard transmission rate $\lambda$ for the edge
$(k,l)$; similarly, in reference~\cite{Olinky04}, Ronen Olinky et
al. also studied the effectiveness of the transmission rate, and in
their work, the transmission rate is $\lambda A(k)$ where $A(k)$
means the probability that a susceptible node actually acquires the
epidemic through an edge connected an infected node with degree $k$.
In these previous studies, the transmission rate on a given edge is
just treated as a function with degrees of the two connecting nodes,
which will induce the transmission rates of two opposite directions
on the same edge are symmetrical. And, their analytical methods and
results which are based on the stationary state of the SIS model may
be not valid in the SIR model.

Thus, in order to make the transmission rate accord with the
realistic cases much more, we take into account the effects of the
weights of edges and the strengths of nodes which are of great
importance measures in the weighted
networks~\cite{BBVpnas,BBVprle1,BBVprle2}. And indeed, the weight
(or the strength) is one of the most important indications in lots
of real networks, for example, in social networks it can represent
the intimacy between individuals; in the internet the weight can
imply the knowledge of its traffic flow or the bandwidths of
routers~\cite{PVCambridge}; in the world-wide airport networks it
can evaluate the importance of a airport~\cite{LANAmaral,BBVpnas},
and so on~\cite{BBVprle1,BBVprle2}. Particularly, for epidemic
spreading, the weight can indicate the extent of frequency of the
contacting of two nodes in scale-free networks, the larger the
weight is, the more intensively the two nodes communicate, at the
same time, the more possible a susceptible individual will be
infected through the edge where the transmission rate is larger.

On the other hand, in the classical SIR model, each infected
individual can establish contacts with all his/her acquaintances
(neighbors) within one time step, that is to say, each infected
node's infectivity equals its degree. But in the real case, a
individual can't contact all his intimate friends, particularly when
he is a patient. In reference~\cite{TZhou06}, the infectivity is
assumed as a constant $A$, which means each infected individual will
generate $A$ contacts at each time step. Recently, Fu et al.
proposed a piecewise linear infectivity~\cite{FuInfect08}, which
means: if the degree $k$ of a node is small, its infectivity is
$\alpha k$; otherwise its infectivity is $A$ as a saturated value
when $k$ is beyond a constant $A/\alpha$. Both the constant or the
piecewise linear method, the heterogeneous infectivity of nodes with
different degrees is not considered as adequately as possible in
scale-free networks, that is to say there may be some nodes with
different degrees which have the same infectivity, and there will be
a large number of such nodes if the constant $A$ is assigned
irrelevantly or the size of underlying networks is infinite. So, in
order to solve these problems, we introduce the nonlinear
infectivity, namely, a infectivity exponent $\alpha$ will be
introduced to take control of the number of contacts that a infected
node generates within one time step, and the $\alpha$ is between 0
and 1 which is convenient to adjust for different scale-free
networks.

In this paper, we present the modified SIR model where the
infectivity exponent $\alpha$ and the weight exponent $\beta$ are
added; based on the modified model, the dynamical differential
equations for epidemic spreading is proposed. We parse the equations
to investigate the threshold behavior and propagation behavior for
epidemic spreading; and the analytical results we obtain is verified
by the necessary numerical stimulations. We show that one can adjust
the values of $\alpha$ and $\beta$ to rebuild the epidemic threshold
to a nonzero finite value for different networks, which can prohibit
or delay the epidemic outbreaks to some extent. And we find $\alpha$
is more sensitive than $\beta$ in the transformation of the epidemic
threshold and epidemic prevalence, which indicates the intrinsic
factor (the infectivity exponent $\alpha$) take more responsibility
than the extrinsic factor (the weight exponent $\beta$) for the
epidemic outbreaks in large scale-free networks.

\section{Standard SIR model}
Epidemic modeling has a history of researching, and mathematicians
also put forward many epidemic
models~\cite{Bailey,May1992,Diekmann}. In the domain of complex
networks, SIR model~\cite{RMMAYSIR01,LloydSCIENCE,MPV02,Moreno03} is
one of the most investigated and classical epidemic models. In the
standard SIR model, individuals can be divided into three classes
depending on their states: susceptible (healthy), infected and
removed (immunized or dead). In order to take into account the
heterogeneity induced by the presence of nodes with different
degrees, we use $S_{k}(t)$, $I_{k}(t)$, $R_{k}(t)$ to denote the
densities of susceptible, infected and removed individuals with
degree $k$ at time $t$, respectively. And these variables are
connected by means of the normalization:
$S_{k}(t)+I_{k}(t)+R_{k}(t)=1$. The global quantities such as the
(average) epidemic prevalence are therefore expressed by an average
over the various degree classes, i.e., $R(t)=\sum_{k}P(k)R_{k}(t)$.
For the standard SIR model, the epidemic evolves by the following
rules: at each time step, a susceptible individual acquires the
infection at the transmission rate $\lambda$ in one contact with any
neighboring infected individual, which means if a susceptible
individual has a edge connecting a infected individual, the disease
will transmitted to the susceptible one through the edge with a
specific probability $\lambda$. On the other hand, the infected ones
will recover and become immune (can't be infected any more) with
rate $\mu$, one can set $\mu=1$ without loss of generality.

For a comparison, firstly we review some classical results from
reference~\cite{MPV02}, where Moreno et al. used the mean field
theory to describe the dynamical differential equations of SIR model
as follows:
\begin{equation}\label{eq1}
\frac{dS_{k}(t)}{dt}=-\lambda
k(1-I_{k}(t)-R_{k}(t))\sum_{k^{'}}P(k^{'}/k)I_{k^{'}}(t),
\end{equation}
\begin{equation}\label{eq2}
\frac{dI_{k}(t)}{dt}=-I_{k}(t)+\lambda
k(1-I_{k}(t)-R_{k}(t))\sum_{k^{'}}P(k^{'}/k)I_{k^{'}}(t),
\end{equation}
\begin{equation}\label{eq3}
\frac{dR_{k}(t)}{dt}=I_{k}(t),
\end{equation}
where $P(k^{'}/k)$ denotes the conditional probability for a node
with degree $k$ to connect a node with degree $k^{'}$. In the
uncorrelated case, Moreno et al. obtained the epidemic threshold:
$\lambda_{c}=\langle k\rangle/\langle k^{2}\rangle$, which implies
the absence of the epidemic threshold in a wide range of scale-free
networks $(\langle k^{2}\rangle \rightarrow \infty$,
$\lambda_{c}\rightarrow 0)$. This result is a bad message for
epidemic controlling and preventing, since the epidemic will prevail
in many real networks with any nonzero value of transmission rate
$\lambda$.

\section{SIR model in weighted network}
In this section, we will give a detailed investigation about the
modified SIR model into which we introduce the weighted transmission
rate and nonlinear infectivity. The results we obtain might deliver
some useful information for the epidemiology. And for a better
analysis, we firstly describe the general differential equations for
SIR model based on the mean field theory, as follows:
\begin{equation}\label{eq4}
\frac{dS_{k}(t)}{dt}=-k(1-I_{k}(t)-R_{k}(t))\sum_{k^{'}}P(k^{'}/k)I_{k^{'}}(t)\frac{\varphi(k^{'})}{k^{'}}\lambda_{k^{'}k},
\end{equation}
\begin{equation}\label{eq5}
\frac{dI_{k}(t)}{dt}=-\mu I_{k}(t)+
k(1-I_{k}(t)-R_{k}(t))\sum_{k^{'}}P(k^{'}/k)I_{k^{'}}(t)\frac{\varphi(k^{'})}{k^{'}}\lambda_{k^{'}k},
\end{equation}
\begin{equation}\label{eq6}
\frac{dR_{k}(t)}{dt}=\mu I_{k}(t),
\end{equation}
where $S_{k}$, $I_{k}$, $R_{k}$ have the same meaning with the
standard SIR model (see section 2); and $\varphi(k^{'})$,
$\lambda_{k^{'}k}$ denote the infectivity of nodes with degree
$k^{'}$ and the transmission rate from nodes with degree $k^{'}$ to
nodes with degree $k$, respectively.

\subsection{The model}
Different from the previous studies, in this paper, we mainly focus
the SIR model on the weighted networks. Among varieties of weighted
patterns in complex networks, making use of nodes's degrees to
express the weights of edges is very important, namely, the weight
between two nodes with degree $k$ and $k^{'}$ may represent as a
function of their degrees~\cite{BBVpnas,BBVprle1,BBVprle2}, i.e.,
$w_{kk^{'}}=w_{0}(kk^{'})^{\beta},$ where the basic parameter
$w_{0}$ and the exponent $\beta$ depend on the particular complex
networks (e.g., in the \emph{E.coli} matabolic network $\beta$ =
0.5; in the US airport network (USAN) $\beta$ = 0.8~\cite{MSTEPL};
in the scientist collaboration networks (SCN) $\beta$ =
0~\cite{BBVpnas}). Noteworthily, the weight $w_{kk^{'}}$ belongs to
an edge, similarly, a node (with degree $k$) also can be measured by
weights, i.e., the strength of a node (with degree $k$), which can
be obtained by summing the weights of the links that connected to
it, i.e., $N_{k}=k\Sigma_{k^{'}}P(k^{'}/k)w_{kk^{'}},$ where $N_{k}$
is the strength of a node with degree $k$. In this paper, for
simplicity, we focus on uncorrelated (also called non-assortative
mixing) networks where the conditional probability satisfies
$P(k^{'}/k)=k^{'}P(k^{'})/\langle k\rangle $~\cite{NewmanCor02}.
Thus, one can obtain $N_{k}=w_{0}\langle k^{1+\beta}\rangle
k^{1+\beta}/\langle k\rangle $.

Here, for each node with degree $k$ we fixed a total transmission
rate which is given by $\lambda k$, and the transmission rate on the
edge from the $k$-degree node to $k^{'}$-degree node, will be
redistributed by the proportion of the k-degree node's strength that
the edge's weight accounts for, that's to say the $\lambda_{kk^{'}}$
can be defined as follows:
\begin{equation}\label{eq7}
\lambda_{kk^{'}}=\lambda k\frac{w_{kk^{'}}}{N_{k}},
\end{equation}
from which we know the more proportion of $N_{k}$ that the weight
$w_{kk^{'}}$ of an edge accounts for, the more possible the disease
will transmit through the edge. In the uncorrelated case, one can
obtain $\lambda_{kk^{'}}=\lambda k^{'\beta}\langle k\rangle /\langle
k^{1+\beta}\rangle $. Moreover, the reasonable total probability
that a susceptible node with degree $k$ will be infected at time
step $t$ is given by $1-\prod_{\forall k^{'}\in
NIDS(t)}(1-\lambda_{kk^{'}})$, where $NIDS(t)$ denotes the degree
sequence of neighboring infected nodes that connect to the
susceptible node with degree $k$ at time step $t$.

On the other hand, from the general differential equations of SIR
model (equations (\ref{eq4}) - (\ref{eq6})), we know $\varphi(k)$
denotes the infectivity of nodes with degree $k$, and here, in the
present model we define it as follows:
\begin{equation}\label{eq8}
\varphi(k)=k^{\alpha}, 0<\alpha\leq1,
\end{equation}
which is to say, each infected individual can establish contacts
with its $k^{\alpha}$ neighbors within one time step. The exponent
$\alpha$ will dominate the infectivity among nodes with different
degrees. Since $0<\alpha\leq1$, it can be adjusted to make the
contacts fall on a more realistic range. And the node's infectivity
will grow nonlinearly with the increasing degree $k$.

We take the simplified expressions of $\varphi(k)$ and
$\lambda_{kk^{'}}$ into the equations (\ref{eq4}) - (\ref{eq6}) with
$\mu=1$ (without lack of generality), we obtain as follows:
\begin{equation}\label{eq9}
\frac{dS_{k}(t)}{dt}=-\frac{\lambda k^{1+\beta}}{\langle
k^{1+\beta}\rangle }S_{k}(t)\theta(t),
\end{equation}
\begin{equation}\label{eq10}
\frac{dI_{k}(t)}{dt}=-I_{k}(t)+ \frac{\lambda k^{1+\beta}}{\langle
k^{1+\beta}\rangle }S_{k}(t)\theta(t),
\end{equation}
\begin{equation}\label{eq11}
\frac{dR_{k}(t)}{dt}=I_{k}(t),
\end{equation}
where $\theta(t)=\sum_{k}k^{\alpha}P(k)I_{k}(t)$. The above
equations combined with the initial conditions $R_{k}(0)=0,
I_{k}(0)=I_{k}^{0}$, and $S_{k}(0)=1-I_{k}(0)-R_{k}(0)=1-I_{k}^{0}$.
And in the general case $I_{k}^{0}$ is very small, then we can
obtain $S_{k}(0)\simeq 1$. Under this approximation, equation
(\ref{eq9}) can be directly integrated, as follows:
\begin{equation}\label{eq12}
S_{k}(t)=e^{-\frac{\lambda k^{1+\beta}}{\langle
k^{1+\beta}\rangle}\phi(t)},
\end{equation}
where
$\phi(t)=\int_{0}^{t}\theta(t)dt=\sum_{k}k^{\alpha}P(k)R_{k}(t)$,
and in the last equality we have made use of equation (\ref{eq11}).

\subsection{Threshold behavior}
In order to obtain some material results for the epidemic threshold
and the average epidemic prevalence, firstly we compute the time
derivative of the magnitude $\phi$~\cite{MPV02}:
\begin{eqnarray}\label{eq13}
\frac{d\phi(t)}{dt}&=& \sum_{k}k^{\alpha}P(k)[1-R_{k}(t)-S_{k}(t)]\nonumber\\
&=& \langle k^{\alpha}\rangle-\phi(t)-\sum_{k}k^{\alpha}P(k)S_{k}\nonumber\\
&=& \langle
k^{\alpha}\rangle-\phi(t)-\sum_{k}k^{\alpha}P(k)e^{-\frac{\lambda
k^{1+\beta}}{\langle k^{1+\beta}\rangle}\phi(t)}.
\end{eqnarray}
For the general $P(k)$ distribution, equation (\ref{eq13}) can not
be solved in a closed form. However, we can still obtain some useful
results in the steady state of the epidemics. Since in the steady
stage with sufficiently large $t$, we have that
$I_{k}(\infty)=I_{k}=0$ and consequently $\lim_{t\rightarrow\infty}$
d$\phi(t)$/dt=0, then one can get the self-consistent equation for
$\phi$ from equation (\ref{eq13}) as follows:
\begin{equation}\label{eq14}
\phi=\langle
k^{\alpha}\rangle-\sum_{k}k^{\alpha}P(k)e^{-\frac{\lambda
k^{1+\beta}}{\langle k^{1+\beta}\rangle}\phi}.
\end{equation}
The value $\phi$=0 is always a (trivial) solution. Then we compute
the second order derivative of the rhs of equation (\ref{eq14}) for
$\phi$, and note that
\begin{equation}\label{eq15}
\frac{d^{2}}{d\phi^{2}}(\langle
k^{\alpha}\rangle-\sum_{k}k^{\alpha}P(k)e^{-\frac{\lambda
k^{1+\beta}}{\langle k^{1+\beta}\rangle}\phi})<0,
\end{equation}
we can see the rhs of equation (\ref{eq14}) is a convex function,
therefore, a nontrivial solution of equation (\ref{eq14}) exists
only if the condition
\begin{equation}\label{eq16}
\frac{d}{d\phi}(\langle
k^{\alpha}\rangle-\sum_{k}k^{\alpha}P(k)e^{-\frac{\lambda
k^{1+\beta}}{\langle k^{1+\beta}\rangle}\phi})|_{\phi=0}>1,
\end{equation}
can be satisfied. This relation implies
\begin{equation}\label{eq17}
\sum_{k}P(k)\lambda\frac{k^{\alpha+\beta+1}}{\langle
k^{1+\beta}\rangle}=\lambda\frac{\langle
k^{\alpha+\beta+1}\rangle}{\langle k^{1+\beta}\rangle}>1,
\end{equation}
and the above inequation defines the epidemic threshold:
\begin{equation}\label{eq18}
\lambda_{c}=\frac{\langle k^{\beta+1}\rangle}{\langle
k^{\alpha+\beta+1}\rangle}.
\end{equation}
Below which the average epidemic prevalence ($R(t)$) will finally be
approximatively null, and above which it will attain a finite value.
One can see if $\alpha=1, \beta=0$, then $\lambda_{c}=\langle
k\rangle/\langle k^{2}\rangle$, which induces the absence of the
epidemic threshold in a wide range of scale-free
networks~\cite{PVBSnet02}. And if $\alpha+\beta=0$, the threshold
will be a finite value given by $\lambda_{c}=\langle
k^{\beta+1}\rangle/\langle k\rangle=\langle
k^{1-\alpha}\rangle/\langle k\rangle\geq1/\langle k\rangle$;
similarly, if $\alpha+\beta=-1$, one can also get a finite threshold
which is $\lambda_{c}=\langle k^{1+\beta}\rangle=\langle
k^{-\alpha}\rangle\geq\langle 1/k\rangle$.

Furthermore, we consider the epidemic threshold in the case of
general scale-free networks of which the degree distribution is
$P(k)=ck^{-\gamma}, 2<\gamma\leq3$, where $c$ is the normalization
constant. Then, we obtain $\langle k^{\beta+1}\rangle=c(\langle
k_{\rm max}^{\beta+2-\gamma}\rangle -\langle k_{\rm
min}^{\beta+2-\gamma}\rangle)/(\beta+2-\gamma)$ and $\langle
k^{\alpha+\beta+1}\rangle=c(\langle k_{\rm
max}^{\alpha+\beta+2-\gamma}\rangle -\langle k_{\rm
min}^{\alpha+\beta+2-\gamma}\rangle)/(\alpha+\beta+2-\gamma)$, where
$k_{\rm max}$ $(k_{\rm min})$ denotes the largest (smallest) degree
in the underlying networks. Substituting into equation (\ref{eq18}),
one can rehandle the epidemic threshold as follows:
\begin{equation}\label{eq19}
\lambda_{c}=\frac{\alpha+\beta+2-\gamma}{\beta+2-\gamma}\times\frac{k_{\rm
max}^{\beta+2-\gamma}-k_{\rm min}^{\beta+2-\gamma}}{k_{\rm
max}^{\alpha+\beta+2-\gamma}-k_{\rm min}^{\alpha+\beta+2-\gamma}}.
\end{equation}
From equation (\ref{eq19}), one can see that the infinite of the
largest degree ($k_{\rm max}\rightarrow \infty$ or equally
$N\rightarrow \infty$, since $k_{\rm max}\propto N^{1/(\gamma-1)}$
~\cite{CoErAvHa}) will make the epidemic threshold $\lambda_{c}$
tends towards zero if $\gamma<\alpha+\beta+2$; on the other hand, if
$\gamma>\alpha+\beta+2$, the epidemic threshold $\lambda_{c}$ is
approximate to be a finite value, given by $\lambda_{c}=k_{\rm
min}^{-\alpha}(\alpha+\beta+2-\gamma)/(\beta+2-\gamma)$. Thus, the
critical border is $\gamma=\alpha+\beta+2$. Although for most real
networks including the internet~\cite{Faloutsos}, the
www~\cite{AlbertJeong}, the world-wide airport
networks~\cite{BBVpnas} and the scientific collaborations
networks~\cite{NewmanSCN01}, the topology exponent $\gamma$ exists
between 2 and 3, which is incidental to induce the absence of the
epidemic threshold, one can adjust the infectivity exponent $\alpha$
and the weight exponent $\beta$ to restore a nonzero threshold for a
given networks (a fixed value of $\gamma$).

\begin{figure}
\begin{center}
\includegraphics[width=0.376\textwidth]{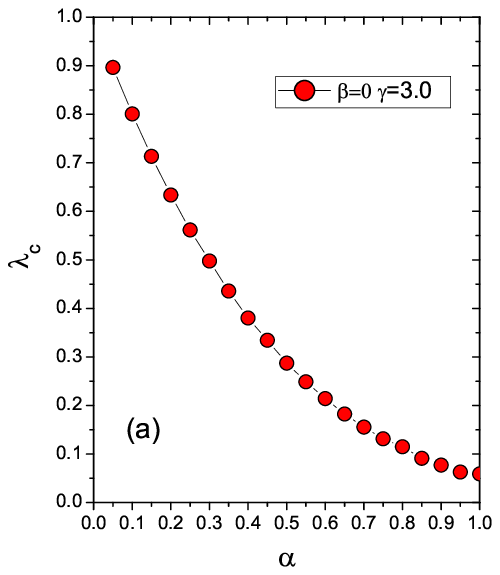}
\includegraphics[width=0.38\textwidth]{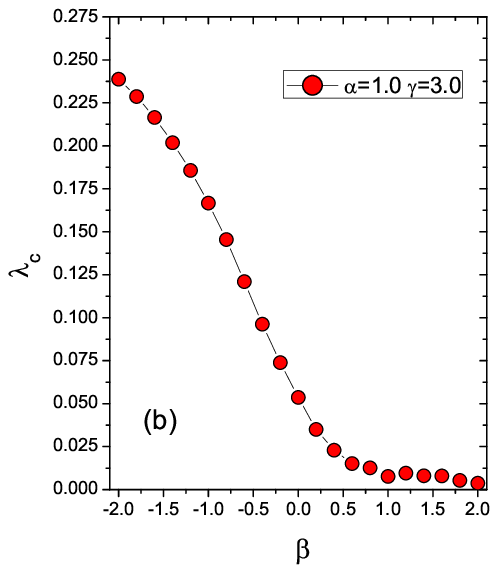}
\caption[kurzform]{\label{threshold12} (a): The epidemic threshold
$\lambda_{c}$ versus $\alpha$ with the exponent $\beta=0$ in BA
networks. (b): The epidemic threshold $\lambda_{c}$ versus $\beta$
with the exponent $\alpha=1.0$ in BA networks.}
\end{center}
\end{figure}

\begin{figure}
\begin{center}
\includegraphics[width=0.61\textwidth]{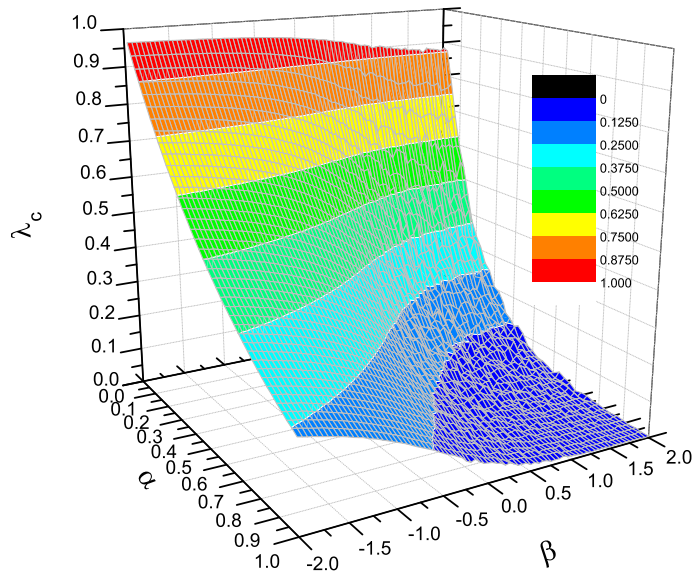}
\caption[kurzform]{\label{threshold3} The epidemic threshold
$\lambda_{c}$ in a 3D-graph, which is made up with various $\alpha$
and $\beta$ in BA networks.}
\end{center}
\end{figure}

In order to get a intuitionistic relation with $\lambda_{c}$,
$\alpha$ and $\beta$, we have performed numerical simulations for
the epidemic threshold. Firstly, on the basis of the simulated
stochastic realizations of SIR model on BA networks~\cite{BA99} of
which the theoretical scale-free exponent is 3 ($\gamma=3$),  we fix
$\beta=0$ and adjust $\alpha$ between 0 and 1 to show the
transformation of epidemic threshold $\lambda_{c}$. In this case the
critical value of $\alpha$ is: $\alpha_{c}=\gamma-\beta-2=1$, below
which $\lambda_{c}$ is a nonzero finite value. As
figure~\ref{threshold12}(a) displays, the value of $\lambda_{c}$ is
greater than 0.01 with the relation $\alpha\ll 1$, vice versa.
Secondly, we fix $\alpha=1.0$ and adjust $\beta$ between -2 and 2 to
show the transformation of $\lambda_{c}$. In this case the critical
value of $\beta$ is: $\beta_{c}=\gamma-\alpha-2=0$, below which
$\lambda_{c}$ is a nonzero finite value. As
figure~\ref{threshold12}(b) displays, at the point of $\beta_{c}=0$,
$\lambda_{c}\simeq0.03$, and $\lambda_{c}$ will take on a much
faster change in $\beta<0$ than the one in $\beta>0$, and the
finiteness of $\lambda_{c}$ is apparent when $\beta<-0.5$. From
figure~\ref{threshold12}, one can see the simulations are consistent
with the analytic results about critical threshold when we consider
the effect of finite scale of the substrate work we use. Moreover,
it is observed that the decreasing trend of $\lambda_{c}$ with
increasing $\alpha$ is much quicker than the one with increasing
$\beta$, that is to say, $\alpha$ is more sensitive than $\beta$ in
the transformation of epidemic threshold $\lambda_{c}$, which means
$\alpha$ is the leading factor for the transformation of
$\lambda_{c}$ in the present model. Figure~\ref{threshold3} displays
the epidemic threshold in a 3D-graph, which is made up with various
$\alpha$ and $\beta$, and the critical condition for a nonzero
finite threshold is $\alpha+\beta<1$. One can see that $\lambda_{c}$
is small in the blue area where the great mass of data meet the
condition that $\alpha+\beta>1$, which is the condition of threshold
vanishing.

\subsection{Propagation behavior}
\begin{figure}
\begin{center}
\includegraphics[width=0.57\textwidth]{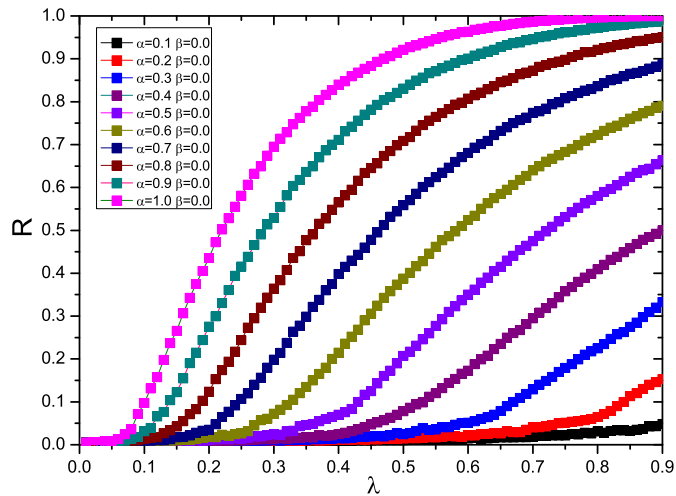}
\caption[kurzform]{\label{rl1} The steady epidemic prevalence $R$
versus $\lambda$ for SIR model in BA networks with $N=10^{4}$,
$\langle k\rangle=6$, $\beta=0$, and $\alpha$=1.0, 0.9, 0.8,
$\cdots$, 0.2, 0.1 (from top to bottom).}
\end{center}
\end{figure}
\begin{figure}
\begin{center}
\includegraphics[width=0.57\textwidth]{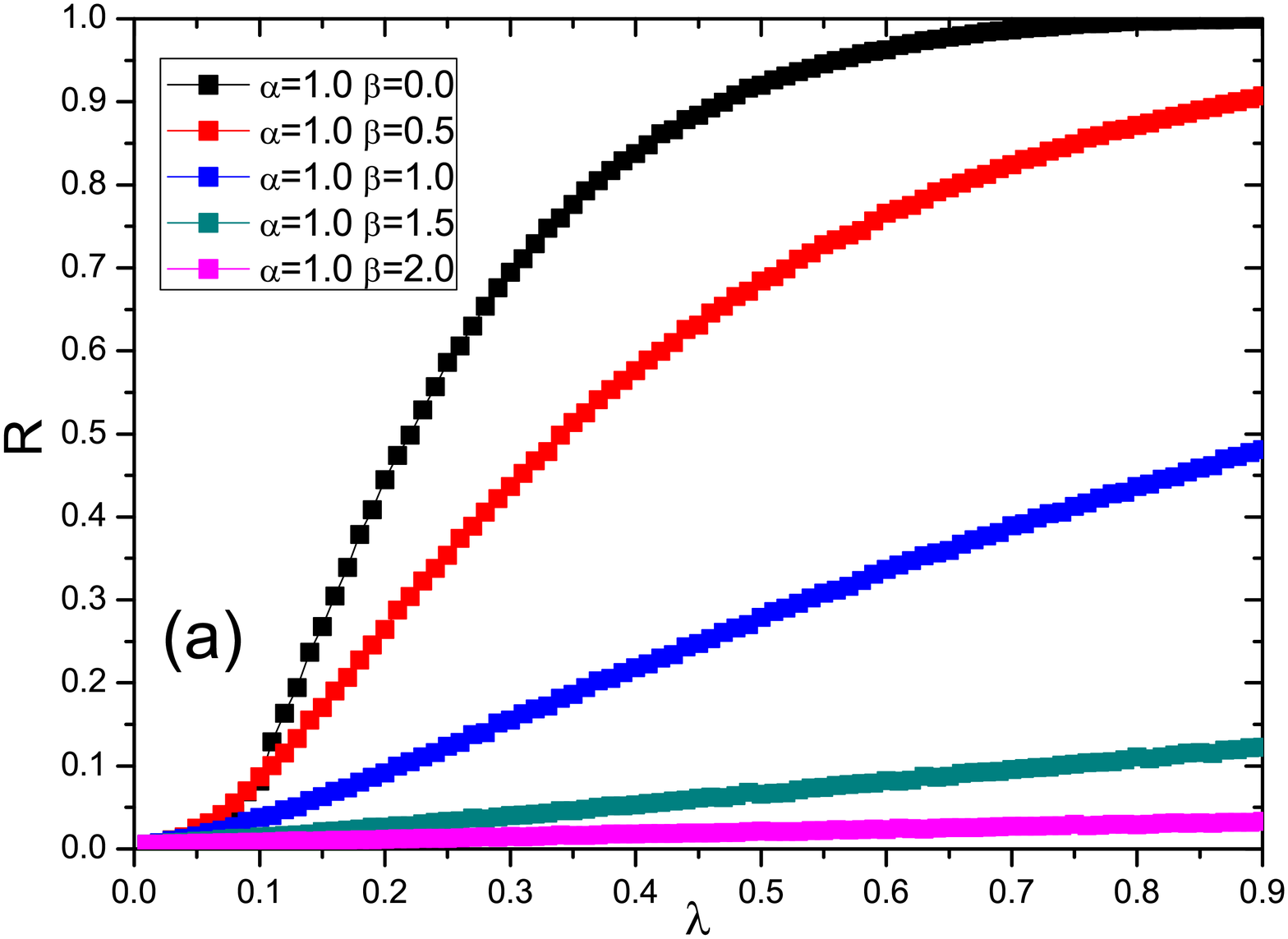}
\includegraphics[width=0.57\textwidth]{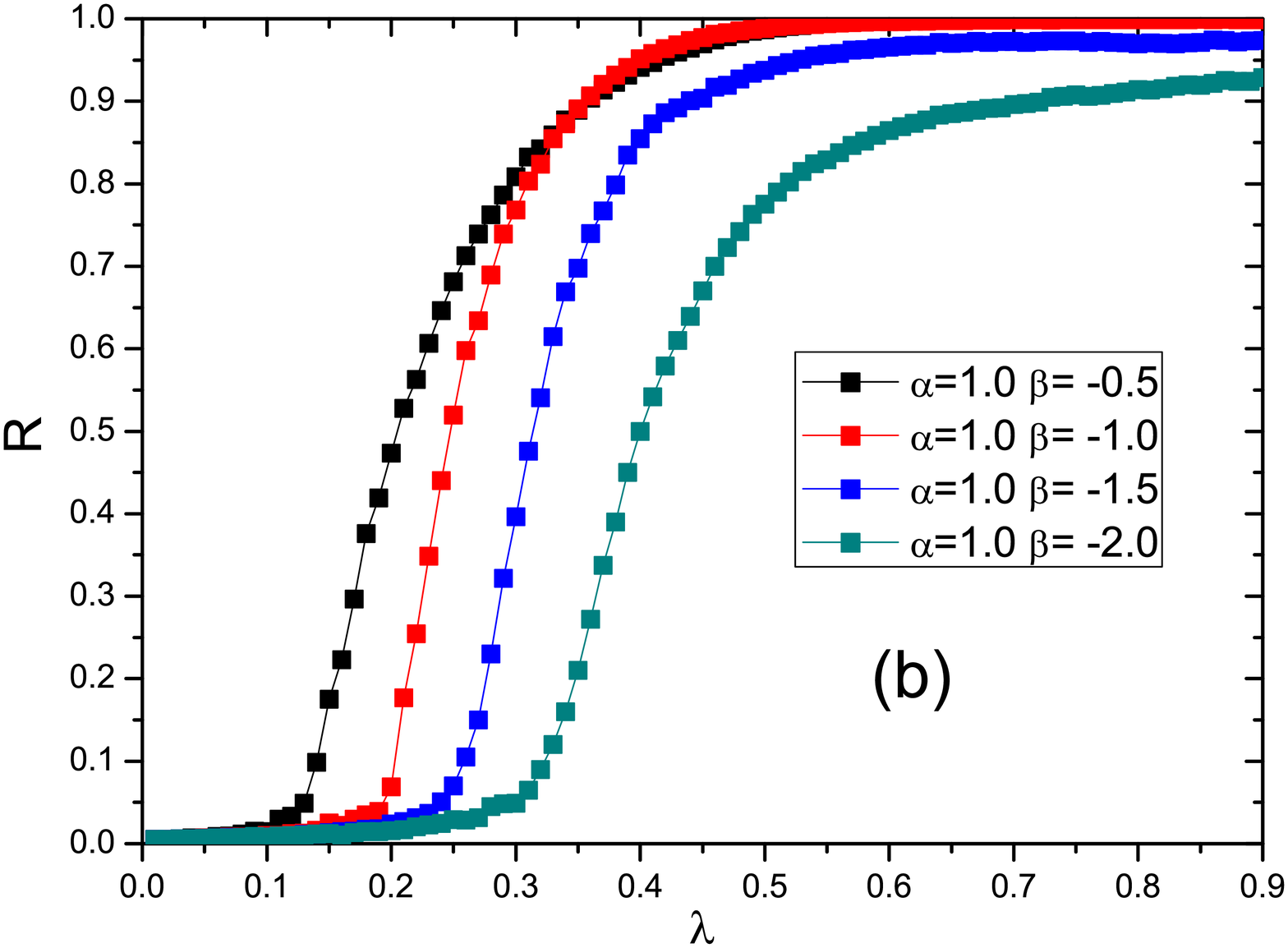}
\caption[kurzform]{\label{rl23} The steady epidemic prevalence $R$
versus $\lambda$ for SIR model in BA networks with $N=10^{4}$,
$\langle k\rangle=6$, $\alpha=1.0$, and (a): $\beta$=0, 0.5, 1.0,
1.5, 2.0 (from top to bottom); (b): $\beta$=-0.5, -1.0, -1.5, -2.0
(from top to bottom).}
\end{center}
\end{figure}
\begin{figure}
\begin{center}
\includegraphics[width=0.57\textwidth]{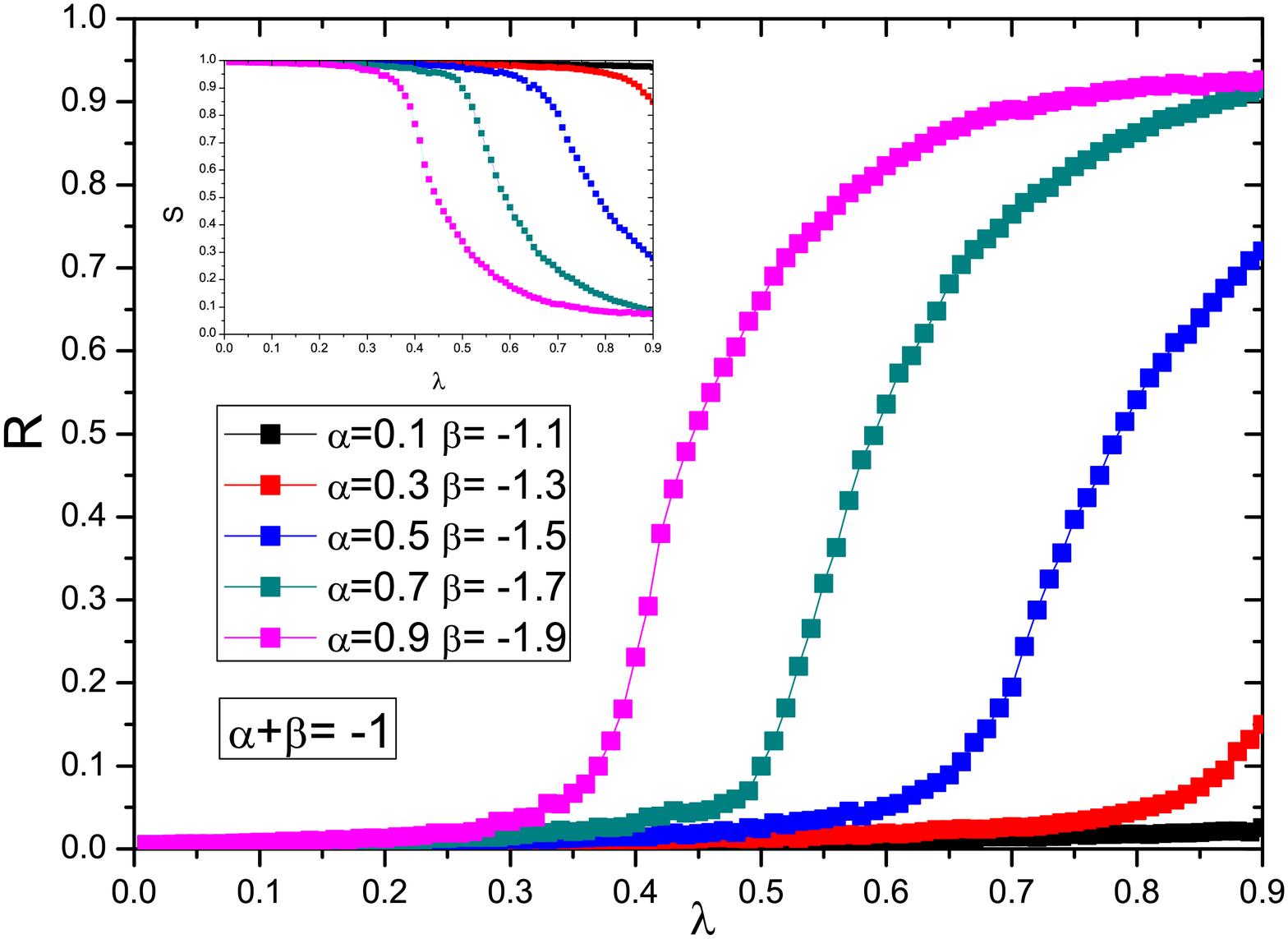}
\caption[kurzform]{\label{rl4} The steady epidemic prevalence $R$
versus $\lambda$ for SIR model in BA networks with $N=10^{4}$,
$\langle k\rangle=6$, $\alpha+\beta=-1$, and from top to bottom:
$\alpha$=0.9, 0.7, 0.5, 0.3, 0.1; accordingly $\beta$=-1.9, -1.7,
-1.5, -1.3, -1.1. The inset shows the susceptible density $S$ versus
$\lambda$ with the same combination of $\alpha$ and $\beta$. }
\end{center}
\end{figure}
\begin{figure}
\begin{center}
\includegraphics[width=0.57\textwidth]{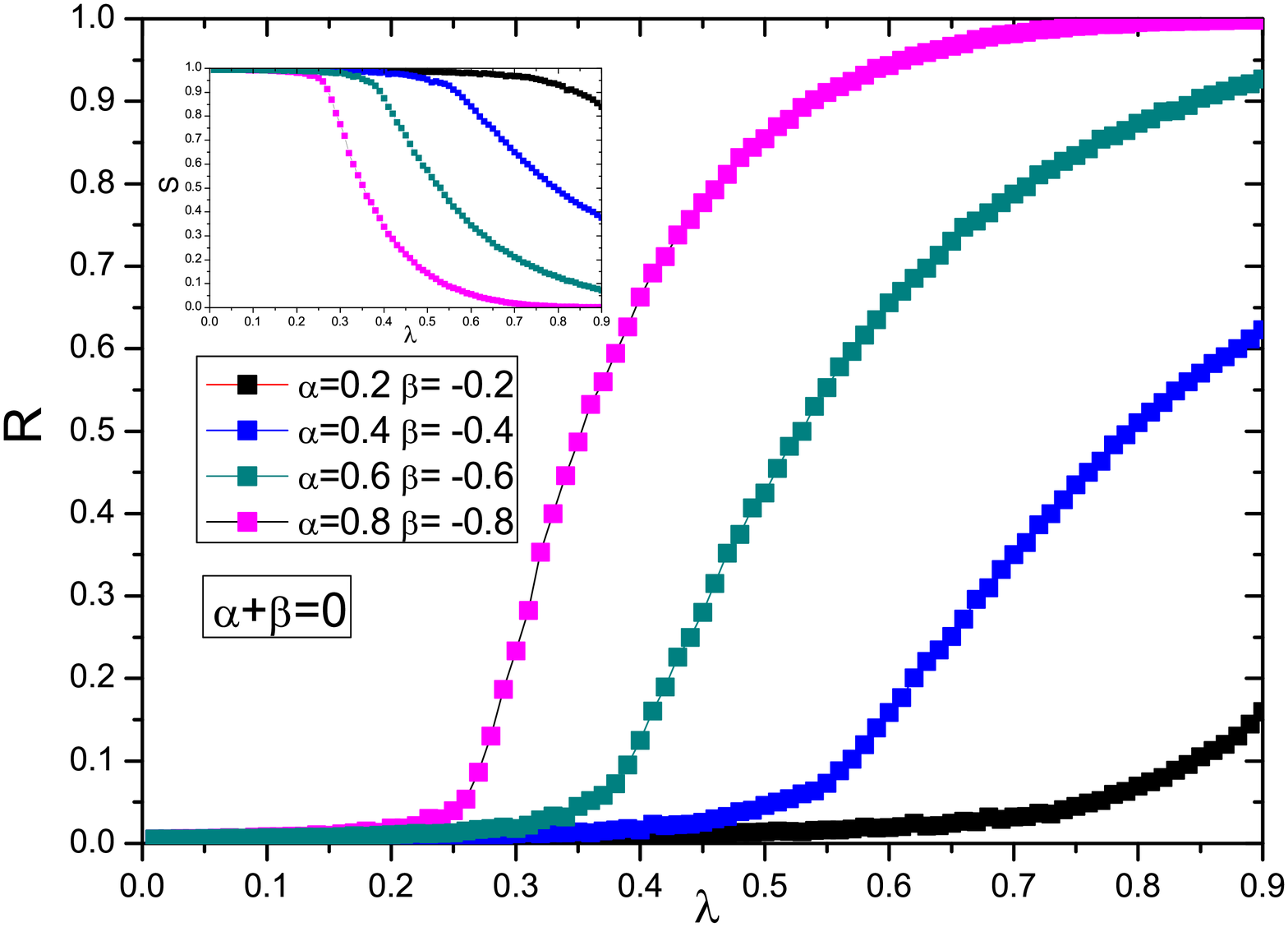}
\caption[kurzform]{\label{rl5} The steady epidemic prevalence $R$
versus $\lambda$ for SIR model in BA networks with $N=10^{4}$,
$\langle k\rangle=6$, $\alpha+\beta=0$, and from top to bottom:
$\alpha$=0.8, 0.6, 0.4, 0.2; accordingly $\beta$=-0.8, -0.6, -0.4,
-0.2. The inset shows the susceptible density $S$ versus $\lambda$
with the same combination of $\alpha$ and $\beta$. }
\end{center}
\end{figure}
\begin{figure}
\begin{center}
\includegraphics[width=0.575\textwidth]{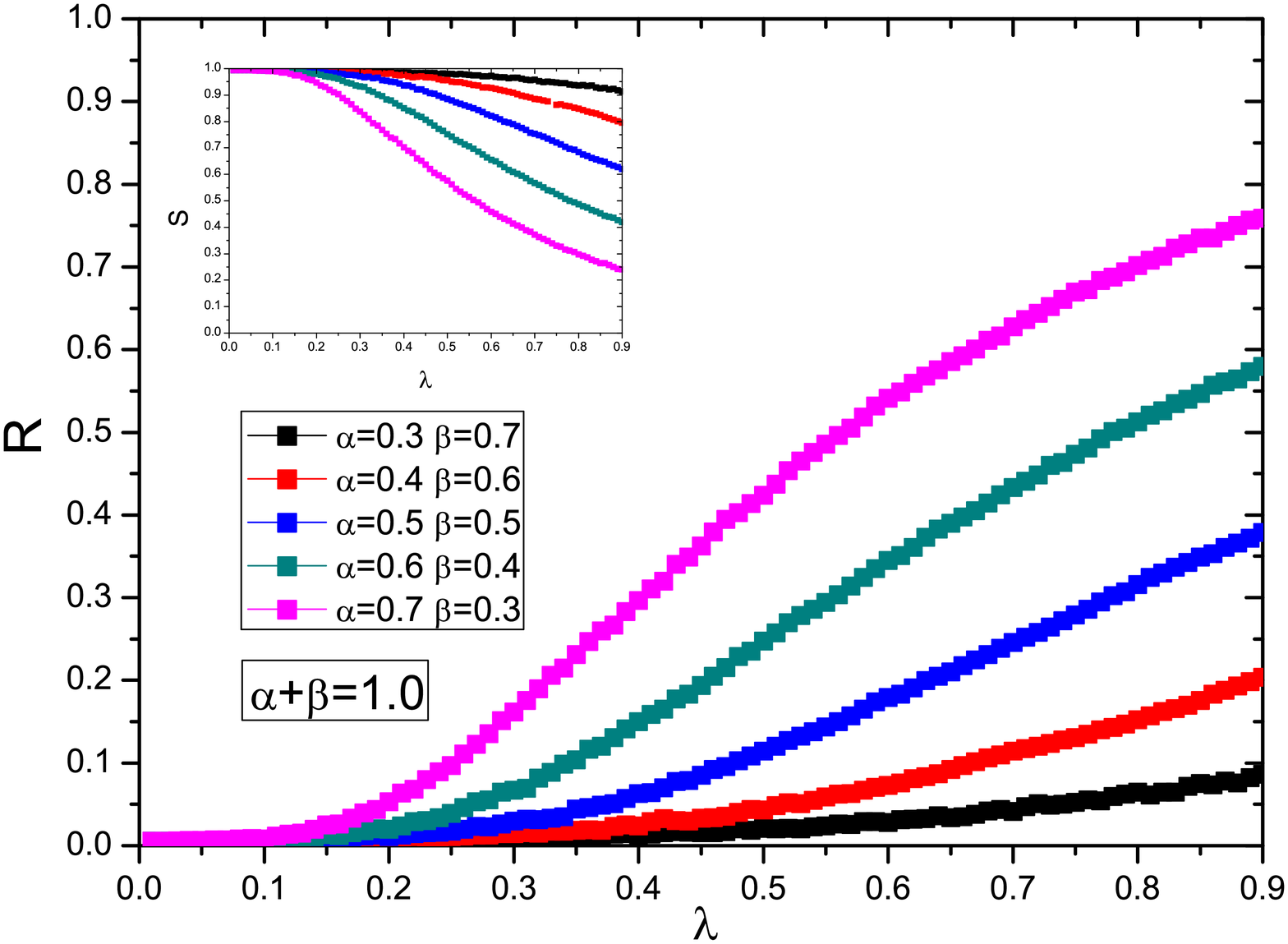}
\caption[kurzform]{\label{rl6} The steady epidemic prevalence $R$
versus $\lambda$ for SIR model in BA networks with $N=10^{4}$,
$\langle k\rangle=6$, $\alpha+\beta=1$, and from top to bottom:
$\alpha$=0.7, 0.6, 0.5, 0.4, 0.3; accordingly $\beta$=0.3, 0.4, 0.5,
0.6, 0.7. The inset shows the susceptible density $S$ versus
$\lambda$ with the same combination of $\alpha$ and $\beta$. }
\end{center}
\end{figure}
\begin{figure}
\begin{center}
\includegraphics[width=0.58\textwidth]{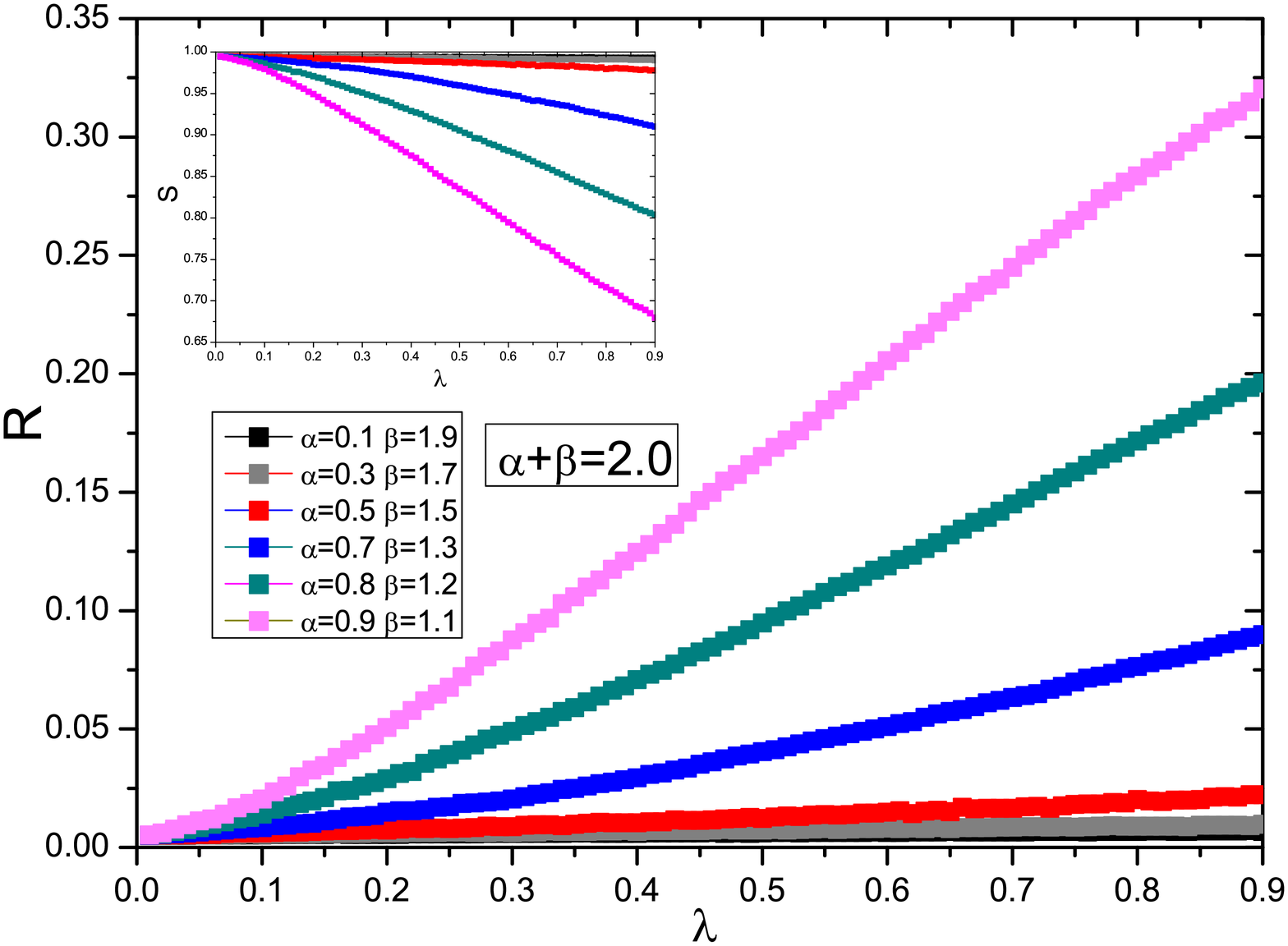}
\caption[kurzform]{\label{rl7} The steady epidemic prevalence $R$
versus $\lambda$ for SIR model in BA networks with $N=10^{4}$,
$\langle k\rangle=6$, $\alpha+\beta=2$, and from top to bottom:
$\alpha$=0.9, 0.8, 0.7, 0.5, 0.3, 0.1; accordingly $\beta$=1.1, 1.2,
1.3, 1.5, 1.7, 1.9. The inset shows the susceptible density $S$
versus $\lambda$ with the same combination of $\alpha$ and $\beta$.
}
\end{center}
\end{figure}
For further investigation of the epidemic dynamics of the present
model, we study the propagation behavior of the epidemic spreading.
Firstly, we investigate the average epidemic prevalence in the
steady stage of epidemic evolution with different combinations of
$\alpha$ and $\beta$. From the analysis in section 3, it is easily
to conclude that $I_{k}=0$ at the epidemic critical point of
$\lambda$, which will induce a quite small $R_{k}$, and we can
approximately get $\phi\simeq0$\ according to the relationship
$\phi= \sum_{k}k^{\alpha}P(k)R_{k}$. Then, expanding the rhs of
equation (\ref{eq14}) for the small $\phi$, and ignoring the
higher-order terms, we obtain
\begin{equation}\label{eq20}
\frac{\langle k^{\alpha+\beta+1}\rangle}{\langle
k^{\beta+1}\rangle}\lambda - \frac{1}{2}\frac{\langle
\alpha+2\beta+2\rangle}{\langle
k^{1+\beta}\rangle^{2}}\lambda^{2}\phi=1.
\end{equation}
Next, we compute the derivative of equation (\ref{eq20}) for
$\lambda$ at the critical point, as follows:
\begin{equation}\label{eq21}
\frac{d\phi}{d\lambda}|_{\lambda_{c}}=\frac{2\langle
k^{\alpha+\beta+1}\rangle^{3}}{\langle k^{1+\beta}\rangle \langle
k^{\alpha+2\beta+2}\rangle}.
\end{equation}
As referred above, one can obtain that
\begin{equation}\label{eq22}
R=\sum_{k}P(k)R_{k}=1-\sum_{k}P(k)e^{-\lambda
k^{1+\beta}\phi/\langle k^{1+\beta}\rangle},
\end{equation}
consequently,
\begin{eqnarray}\label{eq23}
\frac{dR}{d\lambda}|\lambda_{c}&=&\sum_{k}P(k)\lambda_{c}\frac{k^{1+\beta}}{\langle
k^{1+\beta}\rangle}\frac{d\phi}{d\lambda}|\lambda_{c}\nonumber\\
&=&\frac{2\langle k^{\alpha+\beta+1}\rangle^{2}}{\langle
k^{\alpha+2\beta+2}\rangle}.
\end{eqnarray}
Similarly, in the considering of general scale-free networks as
referred above, equation (\ref{eq23}) can be written as follows:
\begin{eqnarray}\label{eq24}
\frac{dR}{d\lambda}|\lambda_{c} &=&
\frac{2c(\alpha+2\beta+3-\gamma)}{(\alpha+\beta+2-\gamma)^{2}}
\frac{(k_{\rm max}^{\alpha+\beta+2-\gamma}-k_{\rm
min}^{\alpha+\beta+2-\gamma})^{2}} {k_{\rm
max}^{\alpha+2\beta+3-\gamma}-k_{\rm min}^{\alpha+2\beta+3-\gamma}}
\propto k_{\rm max}^{\alpha+1-\gamma}.
\end{eqnarray}
The obtained results shows, the exponent $\alpha$ will make a
primary contribution to the velocity of increasing of the steady
epidemic prevalence ($R$) by given the topology of a underlying
network. Combining the analysis in section 3, for a fixed sum of
$\alpha$ and $\beta$, one can conclude that the more ratio of
$\alpha$ to $\beta$ is , the larger the $\lambda_{c}$ is, and the
slowly the $R$ grows as $\lambda$ increases (since the most
large-scale real networks have the relationship
$\alpha+1-\gamma<0$).

For a better understanding of the epidemic propagation behavior, we
take numerical simulations with various combination of $\alpha$ and
$\beta$ on BA networks ($\gamma=3$). Firstly, we investigate the
impact of $\alpha$ and $\beta$ separately, which are the two
particular cases: $\alpha=1.0$ with different $\beta$, and $\beta=0$
with different $\alpha$. Figure~\ref{rl1} displays the effects of
$\alpha$ on the steady epidemic prevalence $R$ with $\beta=0$. As
the figure shows, for a same $\lambda$, one can observe the slope of
$R$ grows as $\alpha$ increases (see $\alpha$=0.9, 0.8, $\cdots$,
0.2, 0.1), which is consistent with the analytical results from
equation (\ref{eq24}). Figure~\ref{rl23} displays $R$ versus
$\lambda$ in the case of $\alpha=1.0$, and figure~\ref{rl23}(a):
$\beta$=0, 0.5, 1.0, 1.5, 2.5 (from top to bottom);
figure~\ref{rl23}(b): $\beta$=-0.5, -1.0, -1.5, -2.5 (from top to
bottom). As shown in figure~\ref{rl23}, the larger the absolute
value of $\beta$ is, the more slowly the $R$ grows. On the other
hand, for a general case ($\alpha\neq1$ and $\beta\neq0$), according
to the critical equation $\alpha+\beta+2=\gamma$ and the algebraic
sign of the sum $\alpha+\beta$, we consider four representative
combinations of $\alpha$ and $\beta$, which are $\alpha+\beta=-1$,
$\alpha+\beta=0$, $\alpha+\beta=1$ and $\alpha+\beta=2$. In each
combination, there also has been divided into several different
configurations by the ratio of $\alpha$ to $\beta$. As shown in
figures (\ref{rl4} - \ref{rl7}), for a fixed sum of $\alpha$ and
$\beta$, one can see the $R$ grows more quickly as the ratio
increasing, which is consistent with our analytical results that
$\alpha$ is a more sensitive factor to $R$. And further more, since
$\alpha+\beta>1$, one can see the epidemic threshold $\lambda_{c}$
is quite small (tends to zero, considering the effects of finite
size) in figure~\ref{rl23}(a) and figure~\ref{rl7}, which is also
consistent with the critical condition of $\lambda_{c}$; otherwise,
for $\alpha+\beta<1$, $\lambda_{c}$ becomes to be a nonzero finite
value as shown in figure~\ref{rl1}, figure~\ref{rl23}(b),
figure~\ref{rl4} and figure~\ref{rl5}. For the early stage of
$\lambda
>\lambda_{c}$ in figures (\ref{rl1} - \ref{rl7}), one can see the $R$ grows in an
exponential form with $\lambda$ increases, then the growth rate will
take off slowly for $\lambda\gg\lambda_{c}$, at last, it will tend
to be zero, which means the value of $R$ have attained a steady
value.

\begin{figure}
\begin{center}
\includegraphics[width=0.56\textwidth]{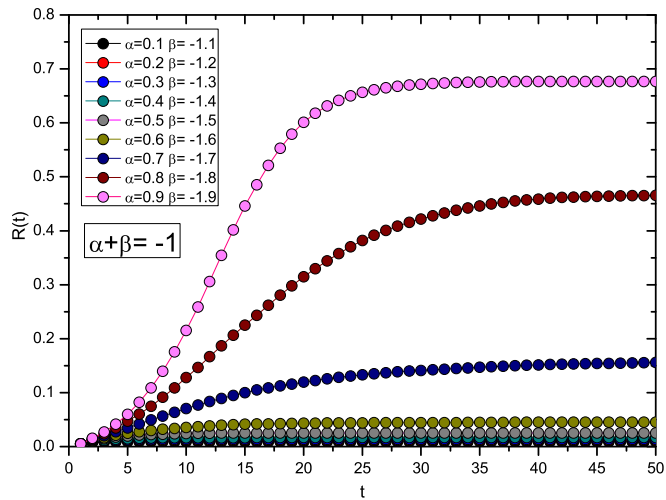}
\caption[kurzform]{\label{rt1} The temporal epidemic prevalence
$R(t)$ versus $\lambda$ for SIR model in BA networks with
$N=10^{4}$, $\langle k\rangle=6$, $\alpha+\beta=-1$, and from top to
bottom: $\alpha$=0.9, 0.8, 0.7, $\cdots$, 0.2, 0.1; accordingly
$\beta$=-1.9, -1.8, -1.7, $\cdots$, -1.2, -1.1.}
\end{center}
\end{figure}

\begin{figure}
\begin{center}
\includegraphics[width=0.56\textwidth]{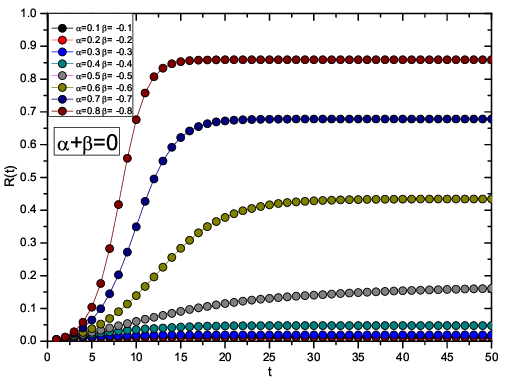}
\caption[kurzform]{\label{rt2} The temporal epidemic prevalence
$R(t)$ versus $\lambda$ for SIR model in BA networks with
$N=10^{4}$, $\langle k\rangle=6$, $\alpha+\beta=0$, and from top to
bottom: $\alpha$=0.8, 0.7, $\cdots$, 0.2, 0.1; accordingly
$\beta$=-0.8, -0.7, $\cdots$, -0.2, -0.1.}
\end{center}
\end{figure}

\begin{figure}
\begin{center}
\includegraphics[width=0.56\textwidth]{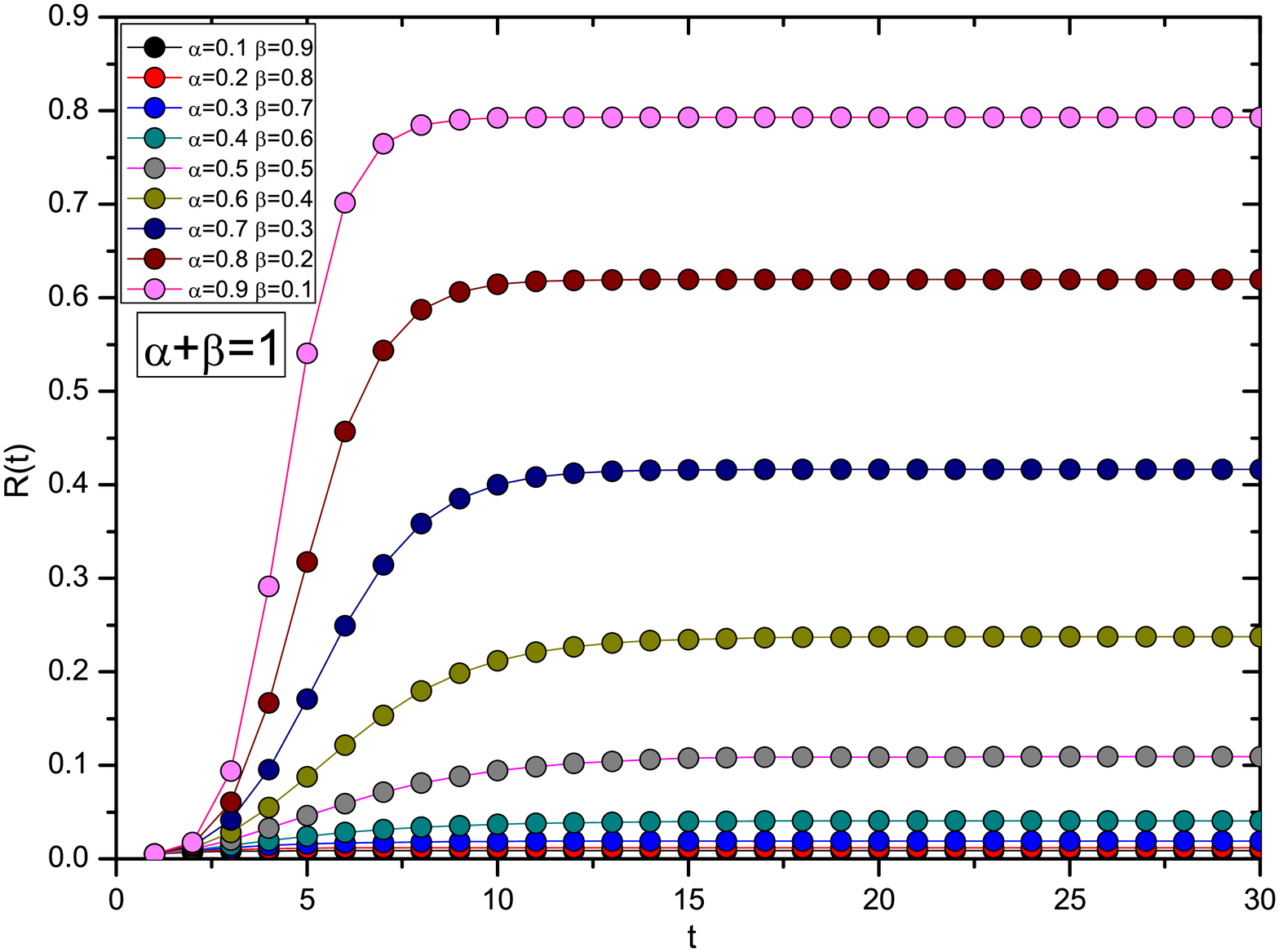}
\caption[kurzform]{\label{rt3} The temporal epidemic prevalence
$R(t)$ versus $\lambda$ for SIR model in BA networks with
$N=10^{4}$, $\langle k\rangle=6$, $\alpha+\beta=1$, and from top to
bottom: $\alpha$=0.9, 0.8, 0.7, $\cdots$, 0.2, 0.1; accordingly
$\beta$=0.1, 0.2, 0.3, $\cdots$, 0.8, 0.9.}
\end{center}
\end{figure}

\begin{figure}
\begin{center}
\includegraphics[width=0.56\textwidth]{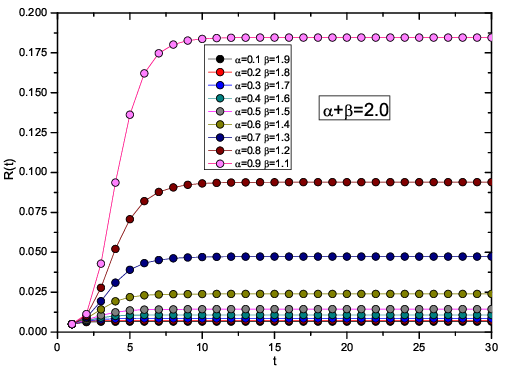}
\caption[kurzform]{\label{rt4} The temporal epidemic prevalence
$R(t)$ versus $\lambda$ for SIR model in BA networks with
$N=10^{4}$, $\langle k\rangle=6$, $\alpha+\beta=2$, and from top to
bottom: $\alpha$=0.9, 0.8, 0.7, $\cdots$, 0.2, 0.1; accordingly
$\beta$=1.1, 1.2, 1.3, $\cdots$, 1.8, 1.9.}
\end{center}
\end{figure}

For investigating the temporal propagation behavior, we simulate the
time behavior of $R(t)$ for SIR model on BA networks with
$\lambda=0.5$. As displayed in figure~\ref{rt1} ($\alpha+\beta=-1$),
figure~\ref{rt2} ($\alpha+\beta=0$), figure~\ref{rt3}
($\alpha+\beta=1$) and figure~\ref{rt4} ($\alpha+\beta=2$), one can
see that, the prevalence $R(t)$ grows in an exponential form in the
early stage, and then stabilizes in a nonzero value as time goes on.
Since $\lambda$ may be smaller than $\lambda_{c}$ when the value of
$\alpha$ is much small such as $\alpha=0.1, 0.2$, thus if
$\lambda<\lambda_{c}$, the steady value of $R(t)$ (i.e., $R$) will
be quite small, which approximatively equals to the initial density
of infected nodes; when $\lambda>\lambda_{c}$, the prevalence $R(t)$
is higher as the ratio of $\alpha$ to $\beta$ gets larger at the
same time of $t$, as the figures display. Moreover, it is observed
that the steady value of $R(t)$ is smaller in figure~\ref{rt4}
compared with the other figures (\ref{rt1} - \ref{rt3}). Although
the exact solution of $R(t)$ about $\alpha$ and $\beta$ which can
demonstrate the difference well is difficult to be managed here,
from a qualitative perspective, we believe that's because of the
trait of SIR model. In figure~\ref{rt4}, $\alpha+\beta=2$ which will
induce a small threshold $\lambda_{c}$, thus at the early stage of
evolution many susceptible nodes will be infected in view of the
relation of $\lambda\gg\lambda_{c}$, and due to the immune rate we
set is unity ($\mu=1$), the old infected nodes will become removed
ones that cant't be infected any more at the same time. Consequently
the optional objects for a infected node will decrease as time
evolutes, Moreover, there will be many infected nodes surrounded by
the removed ones. Under this situation, the epidemic spreading will
arrive at a equilibrium much more quickly, and thus the epidemic
prevalence in the steady stage also will be a small value as
figure~\ref{rt4} displays.

\section{Conclusion and discussion}
To sum up, in this paper, we have investigated the dynamical
behavior of SIR model with weighted transmission rate and nonlinear
infectivity, we present that one can adjust the exponent $\alpha$
and $\beta$ to control the epidemic threshold which is absent for
the standard SIR model in scale-free networks. The critical value
just depends on the exponent $\alpha$ and $\beta$ for a given
topology of networks (a fixed value of $\gamma$), and $\alpha$ is
more sensitive than $\beta$ for the transformation of the epidemic
threshold and epidemic prevalence, which agrees with the numerical
simulations very well. And the numerical results of the time
behavior of $R(t)$ also have been presented, where the remarkable
result is that, for a fixed $\lambda$, the smaller threshold
$\lambda_{c}$ will induce a smaller epidemic prevalence at the
equilibrium.

In a way, epidemic spreading can be managed as a reaction-diffusion
process~\cite{rdcusf,bosonic,rdum}, which also has a very close
relation with information retrieval, peer-trust and influence
spreading. The efficient diffusion or inefficient diffusion maybe
has its merits in various natural and artificial networks. Our work
might deliver some useful information or new insights in designing
data layout, city layout and network layout for performing their
best advantages.

\section*{Acknowledgments}
We benefited from useful discussions with Yichao Zhang, Ming Tang.
This research was supported by the National Basic Research Program
of China under grant No. 2007CB310806, the National Natural Science
Foundation of China under Grant Nos. 60704044, 60873040 and
60873070, Shanghai Leading Academic Discipline Project No. B114, and
the Program for New Century Excellent Talents in University of China
(NCET-06-0376).

\end{document}